# Characterization of Three-Dimensional Microstructures in Single-Crystal Diamond

P. Olivero[1], S. Rubanov[1], P. Reichart[1], B. C. Gibson[1],

S. T. Huntington[1], J. R. Rabeau[1], Andrew D. Greentree[2], J. Salzman[3],

D. Moore[4], D. N. Jamieson[2], S. Prawer[2]

[1] School of Physics, Microanalytical Research Centre, The University of Melbourne, Victoria 3010, Australia.

[2] Centre for Quantum Computer Technology, School of Physics, The University of Melbourne, Victoria 3010, Australia.

[3] Electrical Engineering Department, Israel Institute of Technology, Technion City, Haifa 32000, Israel.

[4] University of Cambridge, Department of Engineering, Trumpington St., Cambridge CB2 1PZ, United Kingdom.

**Abstract**

We report on the Raman and photoluminescence characterization of three-dimensional microstructures created in single-crystal diamond with a Focused Ion Beam (FIB) assisted lift-off technique. The method is based on MeV ion implantation to create a buried etchable layer, followed by FIB patterning and selective etching. In the applications of such microstructures where the properties of high quality single-crystal diamond are most relevant, residual damage after the fabrication process represents a critical technological issue. The results of Raman and photoluminescence characterization indicate that the partial distortion of the $sp^3$-bonded lattice and the formation of isolated point defects are effectively removed after thermal annealing, leaving low amounts of residual damage in the final structures. Three-dimensional microstructures in single-crystal diamond offer a large range of applications, such as quantum optics devices and fully integrated opto-mechanical assemblies.

## 1. Introduction

Because of its extreme physical properties, diamond is a very attractive material for micromachining. The range of applications includes scanning probe tips [1, 2], micro- and nano-electromechanical devices (MEMS, NEMS) [3, 4] and micro-fluidics [5]. In optics, it has found applications in micro-devices [6, 7] and high power laser technologies [8]. Diamond also exhibits a vast inventory of luminescent centers [9]: some of these are very suitable for single photon sources [10, 11] and quantum information processing [12, 13]. However, the fabrication of micro-devices to exploit such attributes is technologically challenging. Diamond microstructures have been fabricated with different techniques based on selective growth/ablation of CVD diamond films with masking and moulding methods [1, 2, 4, 5]. In addition, laser and Focused Ion Beam (FIB) machining have been employed to pattern microstructures directly into diamond [3, 14, 15, 16].

A FIB-assisted lift-off technique, which can be used to create three-dimensional micron-sized cantilever and waveguide structures in single-crystal diamond, has recently been reported by us [17]. The technique was inspired by the diamond lift-off technique, which was initially developed as a method to remove thin layers from bulk diamond samples [18]. This was subsequently employed for laser-assisted micromachining of CVD diamond in two-dimensional micro-gears [19]. Our technique is based on MeV ion implantation combined with thermal annealing to create a buried "sacrificial" layer at a well-defined depth. Patterned milling with a FIB is used to expose defined regions of the buried layer and selective chemical etching results in subsequent lift-off. A final thermal annealing step is employed to remove residual damage produced in the fabrication. In the present work we review the fabrication process and report on the photoluminescence (PL) and Raman characterization of the microstructures.

**2. FIB-assisted lift-off technique**

The samples used were artificial diamonds produced by Sumitomo. The crystals were cut and polished from large single-crystals which were synthesized under ultra-high pressure and temperature (HPHT). All sides of the sample have 100 face orientation, and the crystal is classified as type Ib; the nitrogen concentration is 10-100 ppm, as reported by the manufacturers.

The samples were implanted with 2 MeV $He^+$ ions on the MP2 microbeam line of the 5U NEC Pelletron accelerator at The University of Melbourne. Regions of $100\times100$ $\mu m^2$ were implanted at a fluence of $1.5\times10^{17}$ ions·$cm^{-2}$ using a raster scanning ion beam which was focused to a micrometer-sized spot, providing optimal homogenous ion fluence. The ion beam current was approximately 9 nA, and the implantation of a $100\times100$ $\mu m^2$ region required around 4.5 min to achieve the desired fluence. The nuclear collisions creating lattice defects occur mainly at the end-of range of the implanted ions [20], leaving the surface intact. In the case of 2 MeV $He^+$ ions in diamond, a highly damaged layer is created at a depth of ~3.5 $\mu m$. Fig. 1 shows the ion-induced damage density profile, calculated with "Transport of ions in matter" (TRIM) Monte Carlo simulation code [21]. In the simulation, the value of the atom displacement energy was set to 50 eV [22, 23], while the damage density profile corresponds to the ion fluence of $1.5\times10^{17}$ ions·$cm^{-2}$.

Following ion irradiation, the implanted regions were patterned with a focused 30 keV Ga$^+$ ion beam. The FIB was employed to mill trenches with sub-micron width, which connected the buried highly damaged layer to the sample surface in shapes defined by the ion beam lithography system. To achieve this conductive carbon tape was attached to the sample surface in close vicinity of the patterned region to eliminate significant deflection or de-focusing of the ion beam due to local charging. The ion current used during the micromachining was ~2.7 nA and the spot had sub-micrometer size. In our experimental conditions, we estimated the machining efficiency in diamond as ~0.1 $\mu m^3 \cdot nC^{-1}$, corresponding to ~3 atoms per incoming ion. Such value is comparable to what reported in silicon, and the diamond FIB-machining technique can be optimized with various methods [24]. Figs. 2a and 2b show SEM images of implanted samples patterned to create cantilever and waveguide structures, respectively. Larger scale structures could be obtained by machining of bulk diamond with high power laser ablation with sub-micron spatial resolution [25], however the FIB technique allows the machining of microstructures with spatial resolution down to a few tens of nanometers.

After the FIB patterning, the samples were thermally annealed at a temperature of 550 °C in air for 1 hour. When the damage density is above a critical threshold, the diamond crystal structure cannot be recovered upon thermal annealing [26, 27]. Thus, after annealing, a heavily damaged layer buried at the end of ion range, converts to a non-diamond phase, while in the other regions the crystal structure is partially restored.

Diamond is characterized by high chemical inertness, while non-diamond phases, including amorphous and graphitic carbon phases, can be selectively etched with different techniques, such as annealing in an oxygen atmosphere [18], wet chemical etching [28, 29] and electrochemical etching [30]. Therefore, after the sacrificial buried layer was exposed to the sample surface through the milled trenches, the samples were etched for ∼1 hour in boiling acid (1:1:1 $H_2SO_4$ / $HClO_4$ / $HNO_3$). Buried non-diamond regions connected to a trench, but otherwise covered by diamond, were etched. This allowed unconnected surface layers to lift off, leaving behind the desired structure, which may include undercut regions. Figs. 2c and 2d show SEM images of the same structures shown in Figs. 2a and 2b, respectively, after the etching step that causes the lift-off of the patterned areas. Free-standing cantilever and waveguide structures were created directly in the bulk crystal.

The samples were completed with a final annealing step to remove the residual damage caused by ion implantation and the FIB-micromachining. The samples were annealed at a temperature of 1100 °C in forming gas (4% $H_2$ in Ar), which was used to prevent high-temperature oxidation of the diamond surface.

**3. Characterization**

The multimoded waveguiding behavior of bridge structures like the one shown in Fig. 2d is reported elsewhere [17]. However residual ion induced damage could degrade the device properties and prevent the effective application of our method in this and other devices where the extreme properties of single-crystalline diamond are most relevant (namely, quantum optical devices and high frequency MEMS and resonators). We therefore performed a detailed investigation of the structural and optical properties of our devices as reported here.

*Raman Spectroscopy*. A Renishaw micro-spectrometer was employed to collect Raman spectra in a confocal geometry, with λ=325 nm excitation light from a He-Cd laser. Such excitation wavelength was chosen in order to avoid the stimulation of a broad NV luminescence background caused by ion implantation a nitrogen rich diamond sample, as experienced with 514 nm excitation. The laser light was focused to a spot diameter ~1 μm, with intensity ~200 kW cm$^{-2}$; the Raman signal was recorded with a multichannel collection system, with spectral resolution ~1 cm$^{-1}$. Fig. 3 shows the Raman spectra collected at different stages of the sample processing. Fig. 3a shows the spectra from the implanted sample, before and after the 550 °C annealing. The Raman spectrum from glassy carbon was acquired under the same experimental conditions, and plotted for comparison. The spectra in Figs. 3b and 3d were taken from the cantilever microstructure, before and after the final 1100 °C annealing, respectively. The spectra were acquired from different regions of the microstructure (i.e. the free-standing layer and the bottom surface after the lift-off), as shown in the inset figure and spectra from the pristine regions were plotted for comparison. The spectra reported in the same graph were normalized to the intensity of the first-order diamond line, and displaced vertically for clarity. The probed depth in the material is ~2 μm, thus ensuring that the measurements from the undercut regions are relevant to only the free standing layer. The first order Raman line of diamond is clearly visible in all spectra, although its position, width and relative intensity with respect to other Raman features varies at different stages of the material processing. No features related to graphitic carbon phases (i.e. the D and G peaks, measured in the test graphitic sample as shown in Fig. 3a) are observed in any spectra. Since the probed depth in our measurements does not reach the highly damaged layer at the end of ion range, we can infer that extended graphitic or graphite-like defects are not measured in the diamond structure for the regions that are damaged below the critical threshold fluence for etchability.

In damaged diamond crystal, the first order Raman line is shifted to lower wavenumbers with respect to the 1332 cm$^{-1}$ position observed in the pristine crystal. The downshift is accompanied by line broadening, which is indicative of a decrease in phonon life-time due to the scattering from the irradiation induced defects. The position and width of the first-order diamond peak at different stages of the material processing were determined with Gaussian fits, and are plotted in Figs. 4a and 4b, respectively. In the implanted material, the 7.7 cm$^{-1}$ wide peak was shifted by ~2.6 cm$^{-1}$ (compared to the 3.5 cm$^{-1}$ wide peak from pristine diamond, centered at 1332 cm$^{-1}$). The spectra from the undercut layer in the processed structure exhibit a narrower line and a reduced downshift. Despite the downshift and widening, the peaks are still highly symmetric after background subtraction, indicating little or no phonon confinement. The above mentioned values compare to what reported in reference [31], where significant downshifts and widening of the first order Raman line (up to values of 34 cm$^{-1}$ and 88 cm$^{-1}$, respectively) are measured as a function of induced damage. According to reference [31], our Raman spectrum from the undercut material after 1100 °C annealing indicate a residual damage density <1×10$^{20}$ vacancies cm$^{-3}$.

As shown in Fig. 3, a pronounced peak at 1635 cm$^{-1}$ is observed only in the ion implanted regions after the initial 550 °C annealing. Its intensity is ~49% of the first order Raman intensity. Such a peak has previously been observed in MeV ion implanted single crystal diamonds and has been attributed to a radiation point defect, the so-called "dumb-bell" (or "split interstitial") defect, consisting of an isolated $sp^2$-bonded carbon pair occupying the position of one $sp^3$-bonded carbon atom in the regular diamond lattice [31]. It has been reported that the position of the peak shifts to lower wavenumbers with increasing damage, ranging from 1635 cm$^{-1}$ to 1617 cm$^{-1}$ in highly damaged samples [32]. The measured position of the peak again indicates a damage density in the shallow implanted regions <1×10$^{20}$ vacancies·cm$^{-3}$. It is worth noting that the peak was not observed in the implanted sample, and it was effectively removed after the final 1100 °C annealing. Such peculiar annealing behavior can be explained by the fact that the peak arises from the local mode frequency of a well-defined point defect. The implanted structure is randomly distorted, while the dumb-bell defect is formed upon thermal annealing at temperatures higher than ~300 °C, and annealed back at temperatures higher than ~600 °C.

While the "dumb-bell" peak is indicative of isolated point defects with a defined structure, the random distortion of the crystalline structure induced by ion implantation is reflected in broad Raman features located below 1400 cm$^{-1}$. If disorder is introduced in a crystal structure, the loss of long-range order causes uncertainty of the crystal momentum and a breakdown of the wave-vector selection rules. Therefore in diamond, besides the zone center ($q$=0) line at 1332 cm$^{-1}$, other vibrational modes can contribute to the first-order Raman scattering. In these conditions, the first order Raman spectrum is directly related to the Vibrational Density of States (VDOS) of diamond, which extends below 1400 cm$^{-1}$ [33], weighted by mode coupling constants between the vibrational band and the excitation light. Thus, the Raman intensity $I$ as a function of frequency shift $\omega$ can be expressed as [34]:

$$I(\omega) = \sum_b C_b \cdot \left(\frac{1}{\omega}\right) \cdot [1+\mu(\omega,T)]g_b(\omega) \qquad (1)$$

where index $b$ refers to different vibrational bands, $C_b$ is the band coupling constant, $g_b(\omega)$ is the band VDOS, and $[1+\mu(\omega, T)]$ is the thermal population defined by the Bose-Einstein distribution: $\mu(\omega, T)=[\exp(h\omega/k_BT)-1]^{-1}$. If the band coupling constants are assumed to be constant, the reduced Raman spectrum $I_R(\omega)$ can be defined as:

$$I_R(\omega) = \frac{I(\omega) \cdot \omega}{1+\dfrac{1}{\exp\left(h\omega/k_BT\right)-1}} \qquad (2)$$

The reduced Raman spectrum is expected to qualitatively match the diamond VSDOS, since the different Raman-scattering coupling constants are not taken into account. Fig. 5 shows the reduced Raman spectra from the implanted sample and after subsequent annealing steps, corresponding to the spectra plotted in Fig. 3. The spectra have been normalized to the intensity of the 1332 cm$^{-1}$ peak, which has not been plotted to full scale in order to improve the visibility of the other features. The diamond VDOS from Ref. 33 is plotted in the inset of Fig. 5 for comparison and the spectrum from the implanted sample qualitatively resembles the theoretical curve, although the fine structures of the VDOS are not exactly reproduced. With subsequent annealing processing, the ordered crystal structure is progressively recovered, restoring the selectivity of the wave-vector rules and thus suppressing all the Raman features below 1400 cm$^{-1}$, with the exception of the zone center line.

In reference [31], the ratio between the step height of the amorphous features at 1245 cm$^{-1}$, with respect to the background intensity above 1400 cm$^{-1}$ and the height of the first order diamond line, was used as a figure of merit (defined as the "amorphous fraction") to estimate the degree of amorphization in the crystal structure. In Fig. 6 the amorphous fraction is plotted at different stages of the material processing, where the initial value of ~63% in the implanted material decreases to ~6% in the final structure. Finally, it is worth noting the high quality of the Raman spectra from the bottom surfaces after the lift-off of the patterned areas. Their first order Raman lines are only very slightly downshifted and widened (see Fig. 4). In addition, the spectra do not exhibit measurable Raman features related to disorder (see Fig. 5), thus leading to very small or null "amorphous fraction" values (see Fig. 6). This can be expected since such regions extend beyond the range of the implanted ions, but it also indicates that any non-diamond carbon phases are very efficiently removed in the etching process. Furthermore, the surfaces are extremely smooth and regular. Surface roughness measurements performed with Atomic Force Microscopy (AFM) confirm that its RMS roughness is ~2 nm, which is comparable to that of the original diamond surface. This fact indicates that in the lift-off process, the damage threshold defines a very sharp interface between the diamond structure and the etchable layer. The ability to create microstructures with smooth surfaces is extremely relevant for photonic-based applications.

*Photoluminescence.* PL spectra were acquired with the same experimental setup used for Raman spectroscopy, using a $\lambda$=514 nm excitation light from an Ar$^+$ laser; the excitation power was focused to a spot diameter ~1 $\mu$m, with intensity ~100 kW cm$^{-2}$. Fig. 7 shows spectra from the induced luminescence from undercut regions, acquired at different stages of the final 1100 °C annealing. Spectra collected at cryogenic temperatures from pristine and undercut regions are also reported for comparison. The PL features characteristic of the diamond $NV^0$ and $NV^-$ centers [9] are clearly visible in the spectrum from the pristine region. In the undercut regions this spectrum is superimposed with fringes that are attributed to the interference of induced PL radiation after multiple internal reflections in the free standing structure, as schematically shown in the inset. It can be observed that the fringe visibility improves as the structure progressively recovers the original optical transmittivity and the etched bottom surface becomes smoother and more reflective.

The thickness of the free standing diamond layer was estimated from the period of the interference pattern at normal incidence, using the formula [35]:

$$d = \frac{1}{2 \cdot n(\lambda) \cdot \left[\left(\frac{1}{\lambda_1}\right) - \left(\frac{1}{\lambda_2}\right)\right]} \quad (3)$$

where $d$ is the layer thickness, $n(\lambda)$ is the refractive index of diamond, and $\lambda_{1,2}$ are the wavelengths corresponding to two consecutive maxima (or minima) in the interference pattern. The resulting thickness is $d = (3.4 \pm 0.1)\,\mu m$. Using this value, and the damage profile plotted in Fig. 1, we estimate that the threshold value upon which the diamond structure is not recovered upon thermal annealing in our samples to be $\sim 9 \times 10^{22}$ vacancies·cm$^{-3}$. It should be noted that the uncertainty in the layer thickness, although relatively small (~3%), determines a large uncertainty in the estimation of a such threshold value. Moreover, the refractive index of pristine diamond was used.

Nonetheless, our estimation of the damage threshold is significantly higher than the value of $1\times10^{22}$ vacancies·$cm^{-3}$ reported by Uzan-Saguy et al. [27]. It is worth noting that, if a threshold value of $1\times10^{22}$ vacancies $cm^{-3}$ was used, the corresponding layer thickness would be significantly lower (~2.8 µm) than the estimations from both interferometry and SEM measurements. We believe this discrepancy can be resolved by considering the differences between the implantations used in the present study and in the cited work. In reference [27] heavy ions at sub-MeV energies were used to create a shallow (<170 nm) damaged layer which was exposed to the sample surface. In our case, the sacrificial layer was deeply buried under a large, relatively intact, diamond cap layer preventing the expansion associated with the conversion to the less dense non-diamond phases. Therefore, the high internal pressure may effectively raise the threshold damage level at which the diamond structure is not recovered upon thermal annealing. Previous works [36, 32] indicate that in MeV ion implantation of diamond, the damage process at the end of ion range is prevented by large internal pressure, and that a buried structure damaged above the critical value of $1\times10^{22}$ vacancies·$cm^{-3}$, can even re-convert entirely to diamond upon thermal annealing.

## 4. Conclusions

The creation of three-dimensional microstructures, such as cantilevers and waveguides, was demonstrated in bulk single crystal diamond with a FIB-assisted lift-off technique. The method presented is based on the following steps:

- 2 MeV $He^+$ ion implantation, to create a buried highly damaged layer;

- patterning the implanted areas with a focused 30 keV $Ga^+$ ion beam, to connect the buried layer with the sample surface in defined shapes with sub-micrometer resolution;
- thermal annealing at 550 °C, to convert the highly damaged regions (i.e. the regions where the damage density exceeds a critical threshold) into etchable non-diamond phase, while partially recovering the pristine structure in less damaged regions;
- wet chemical etching of the sacrificial layer, to lift off unconnected regions, thus leaving undercut (i.e. free-standing) structure in the bulk sample;
- final thermal annealing at 1100 °C, to remove the residual damage from the microstructures.

The damage created during the fabrication process was evidenced by the shift and widening of the first order diamond line, together with the emergence of a sharp Raman peak attributed to isolated defects in the crystal structure. Moreover, the distortion of the lattice determined the onset of broad Raman features due to the partial relaxation of the wave-vector selection rules. The evolution of the above mentioned Raman features in the machining of the microstructures consistently indicates that:

- after the 550 °C thermal annealing, the highly damaged regions are converted to etchable non-diamond phase; such conversion occurs above a well-defined damage threshold, which defines a very sharp interface between diamond and the etchable phase;

- the 1100 °C annealing effectively removes the ion induced damage (consisting in amorphized regions and point defects), although a relatively small amount of residual damage is still present in the final structures;
- the Raman spectra from the bottom surfaces which are left after the lift-off of patterned areas indicate that in the selective etching process any non-diamond phases are effectively removed at the interface with the sacrificial layer.

The results suggest that, for applications where high quality single crystal diamond are needed, further processing strategies might be required to fully remove residual damage from the microstructures. This can be achieved by varying annealing conditions and etching methods. The use of micro-patterned masks could be employed to protect defined areas of the devices, taking advantage of the straggling of ions near the end of range to effectively undercut the masked regions. Also, it is known that damage induced by cold implantation (instead of room temperature as employed here) can be more effectively removed with subsequent annealings.

Interferometry measurements from the photoluminescence spectra allowed the estimation of the thickness of the free-standing layers. When compared with the density profile of damage created by 2 MeV $He^+$ implantation, the estimated damage threshold above which the diamond structure is not recovered upon thermal annealing is estimated to be $\sim 9\times 10^{22}$ vacancies $cm^{-3}$, which is significantly higher than what previously estimated for shallow sub-MeV implantations [27]. This discrepancy is attributed to the large internal pressure applied to the buried damaged layer by the surrounding diamond structure.

In summary, Raman and photoluminescence characterization indicate that the FIB-assisted lift-off technique allows for the creation of three-dimensional microstructures in single crystal diamond with good structural properties and control on the damage induced during the fabrication process. These microstructures offer a large range of applications, where the superior structural and optical properties of single crystal diamond are required, such as quantum optical devices and fully integrated opto-micro-mechanical assemblies.

**Acknowledgements**

This work was supported by the Australian Research Council, the Australian government and by the US National Security Agency (NSA), Advanced Research and Development Activity (ARDA), and the Army Research Office (ARO) under contract number W911NF-05-1-0284 and DARPA QUIST.

B. Gibson is supported by the International Science Linkages programme established under the Australian Government's innovation statement Backing Australia's Ability.

S. Huntington is supported by the Nanostructural Analysis Network Organization major national research facility, funded by the Department of Education, Science and Training, Australian Government.

**Figure captions**

Fig.1: TRIM Monte Carlo simulation of the damage density profile induced in diamond by 2 MeV He$^+$ ions compared with the thickness of the free standing layer evaluated with interferometry measurements. The buried region where the damage density is above the estimated threshold is highlighted.

Fig. 2: SEM images of micro-structures machined in single crystal diamond. Figures 2a and 2b show the FIB-patterned regions for the fabrication of a cantilever and waveguide structure, respectively. The milled linear trenches expose the buried sacrificial layer to the surface in the desired shapes. Figures 2c and 2d show the three-dimensional structures, respectively, in the same samples after the completion of the lift-off step.

Fig. 3: Raman spectra collected at different stages of the sample processing. Fig. 3a shows the Raman spectra from the implanted sample, before and after the first annealing steps, together with a reference spectrum from glassy carbon. Figs. 3b and 3d show spectra from the cantilever microstructure, before and after the 1100 °C annealing, respectively. The spectra were acquired from different regions of the microstructure, as shown in the inset figure. The spectra reported in the same graph are normalized to the intensity of the first-order diamond line, and displaced vertically for clarity.

Fig. 4: Position and width of the first order diamond Raman peak at different stages of the sample processing, obtained from the spectra reported in Fig. 3. As the implanted material is progressively annealed, the Raman peak sharpens and shifts towards the 1332 cm$^{-1}$ position characteristic of the pristine material. The residual downshift and widening after the final 1100 °C annealing indicates that although most of the ion induced damage is effectively removed, some residual damage is still present in the crystal lattice.

Fig. 5: Reduced Raman spectra at different stages of the material processing, corresponding to the spectra shown in Fig. 3. The spectra have been normalized to the intensity of the zone center peak, not plotted in full scale to improve the visibility of the other features. The inset figure shows the diamond VDOS from Ref. 33 for comparison.

Fig. 6: Amorphous fraction at subsequent stages of the material processing. The value of the figure of merit is defined as the ratio between the step height of the amorphous features at 1245 cm$^{-1}$ with respect to the background intensity above 1400 cm$^{-1}$ and the height of the first order diamond line. After the final 1100 °C annealing, the parameter is reduced to about 10% of its initial value.

Fig. 7: Photoluminescence spectra taken from undercut and pristine regions. As the 1100 °C annealing time increases, the intrinsic emission spectrum from free standing regions (a, b, c) gets superimposed with sharp interference fringes that can be attributed to internal reflections, as schematically shown in the inset (note: the excitation laser beam is normal to the sample surface). As a reference, spectra from the pristine and undercut regions (d, e) were acquired at liquid nitrogen temperature and the positions of $NV^0$ and $NV^-$ zero phonon lines are indicated by the arrows.

**Figures**

**Fig. 1**

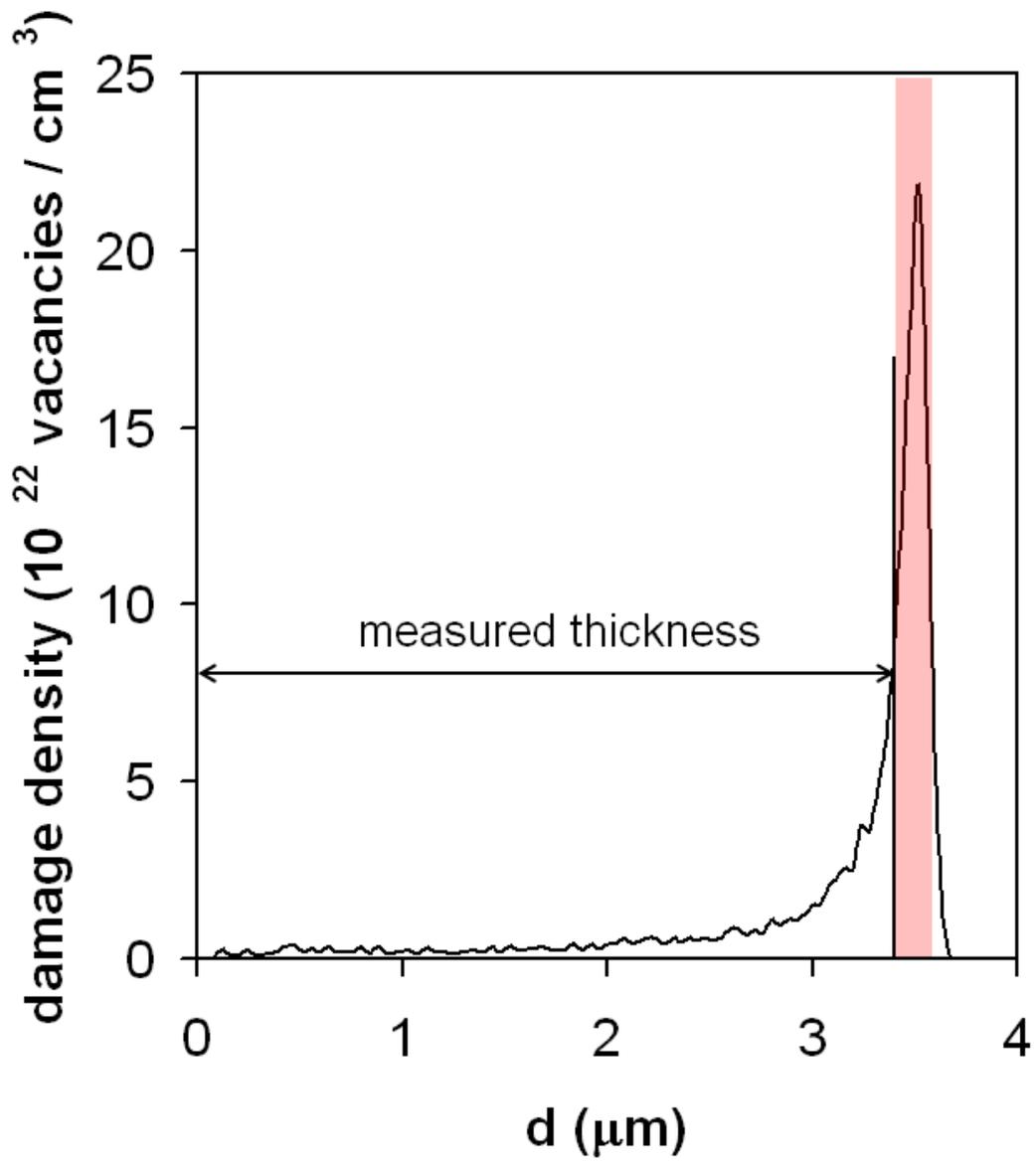

**Fig. 2**

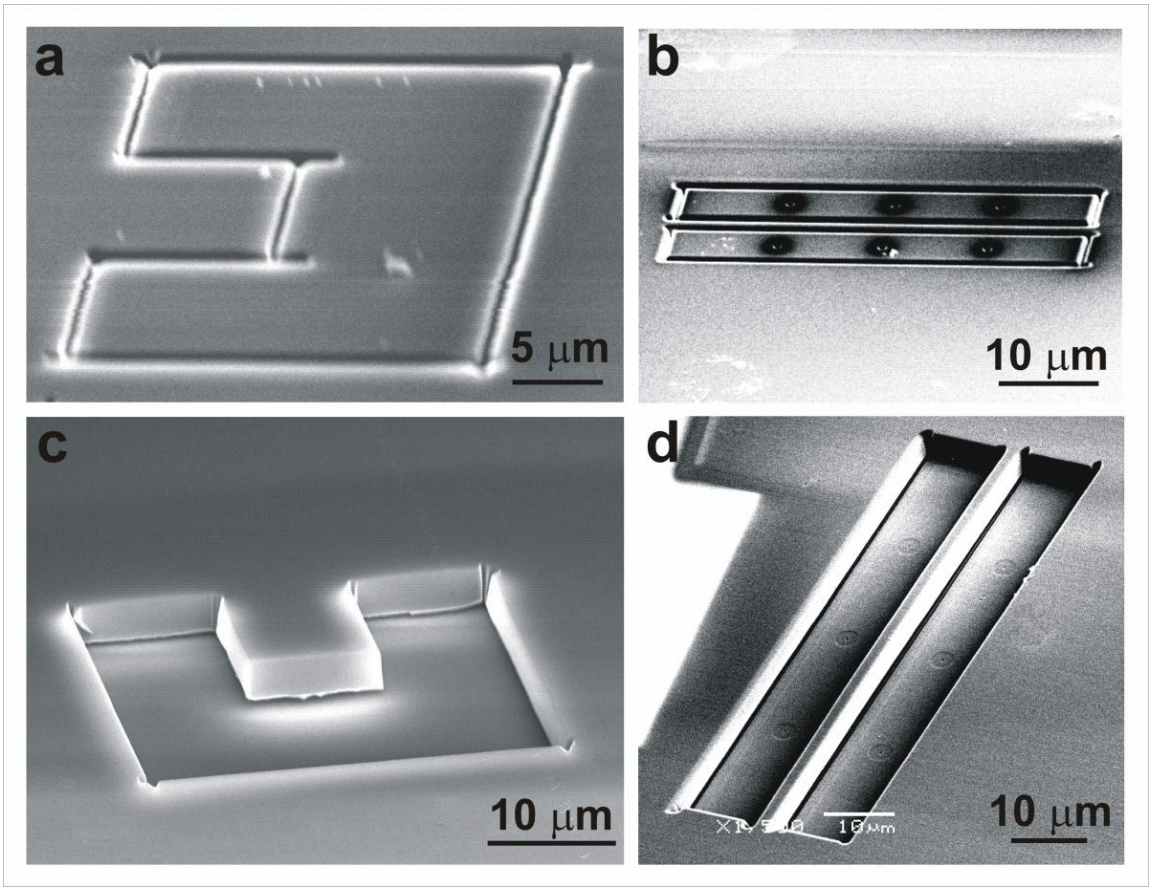

**Fig. 3**

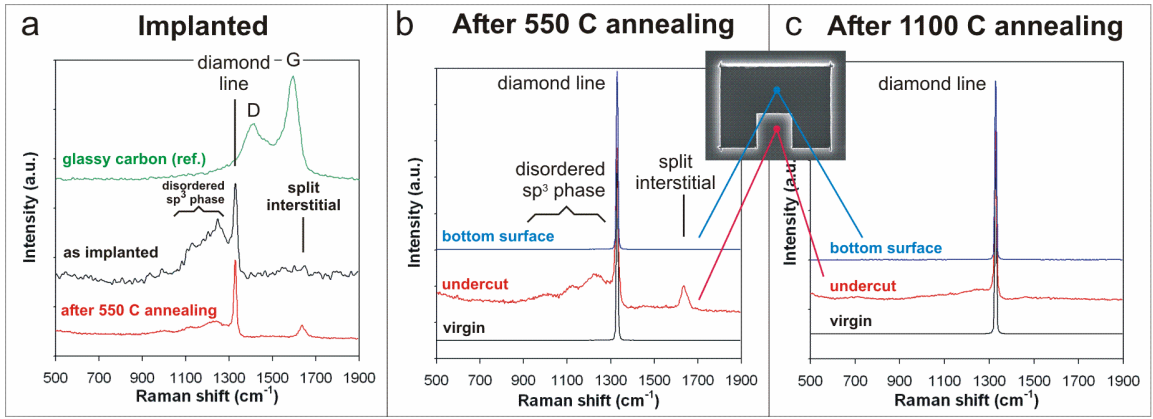

**Fig. 4**

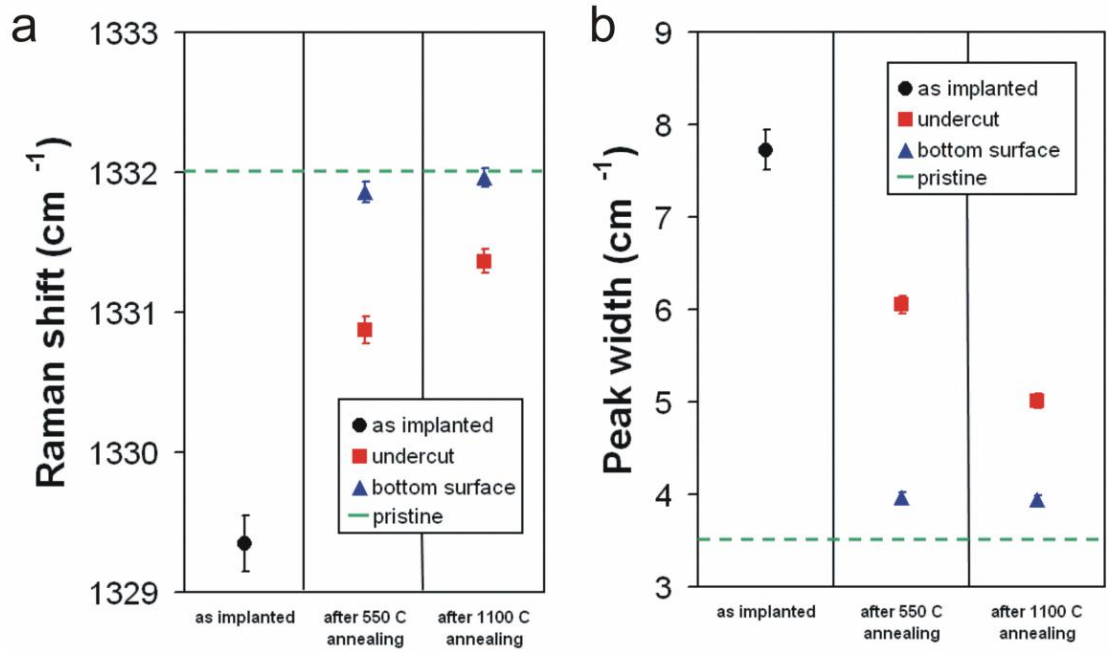

**Fig. 5**

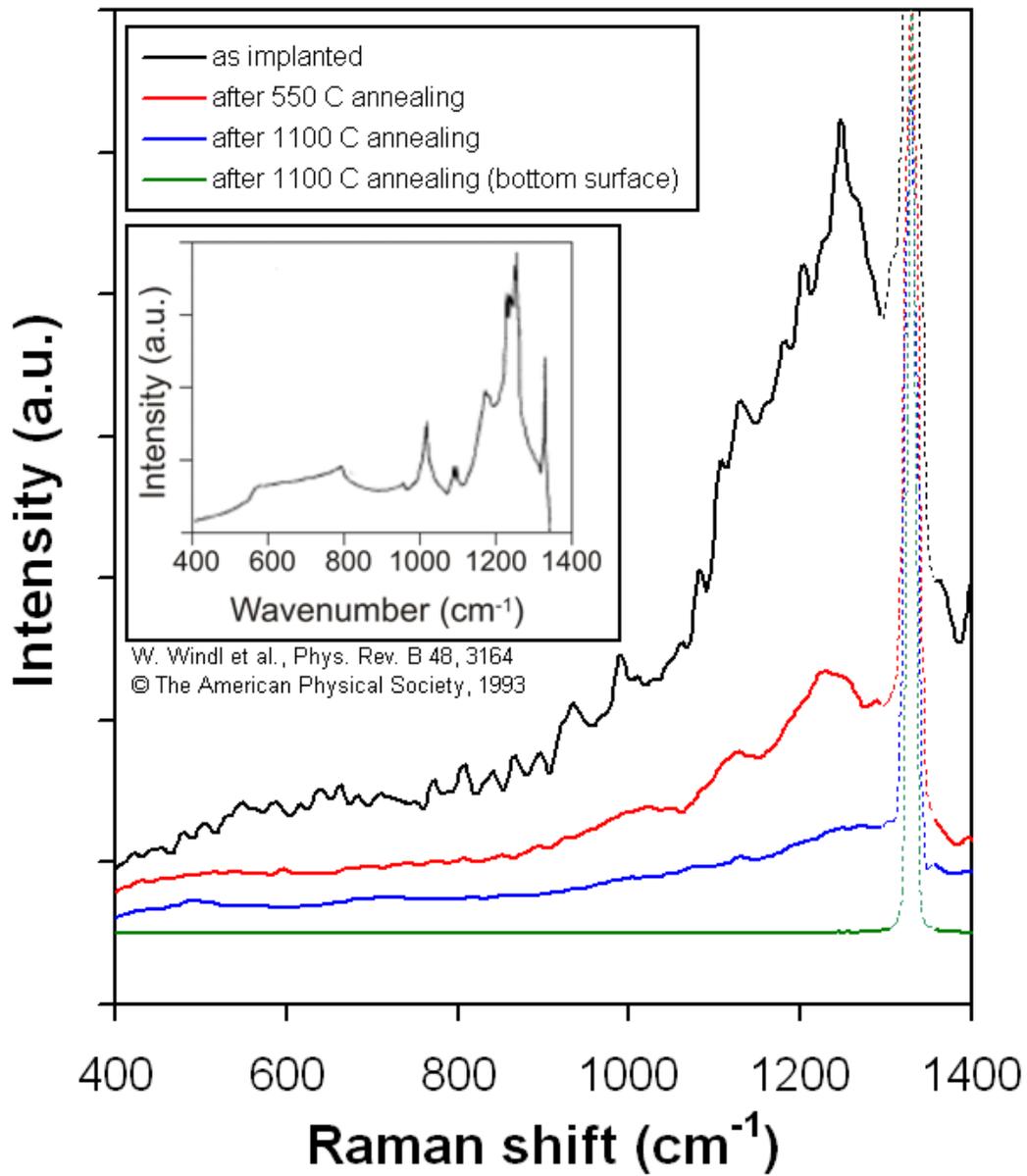

**Fig. 6**

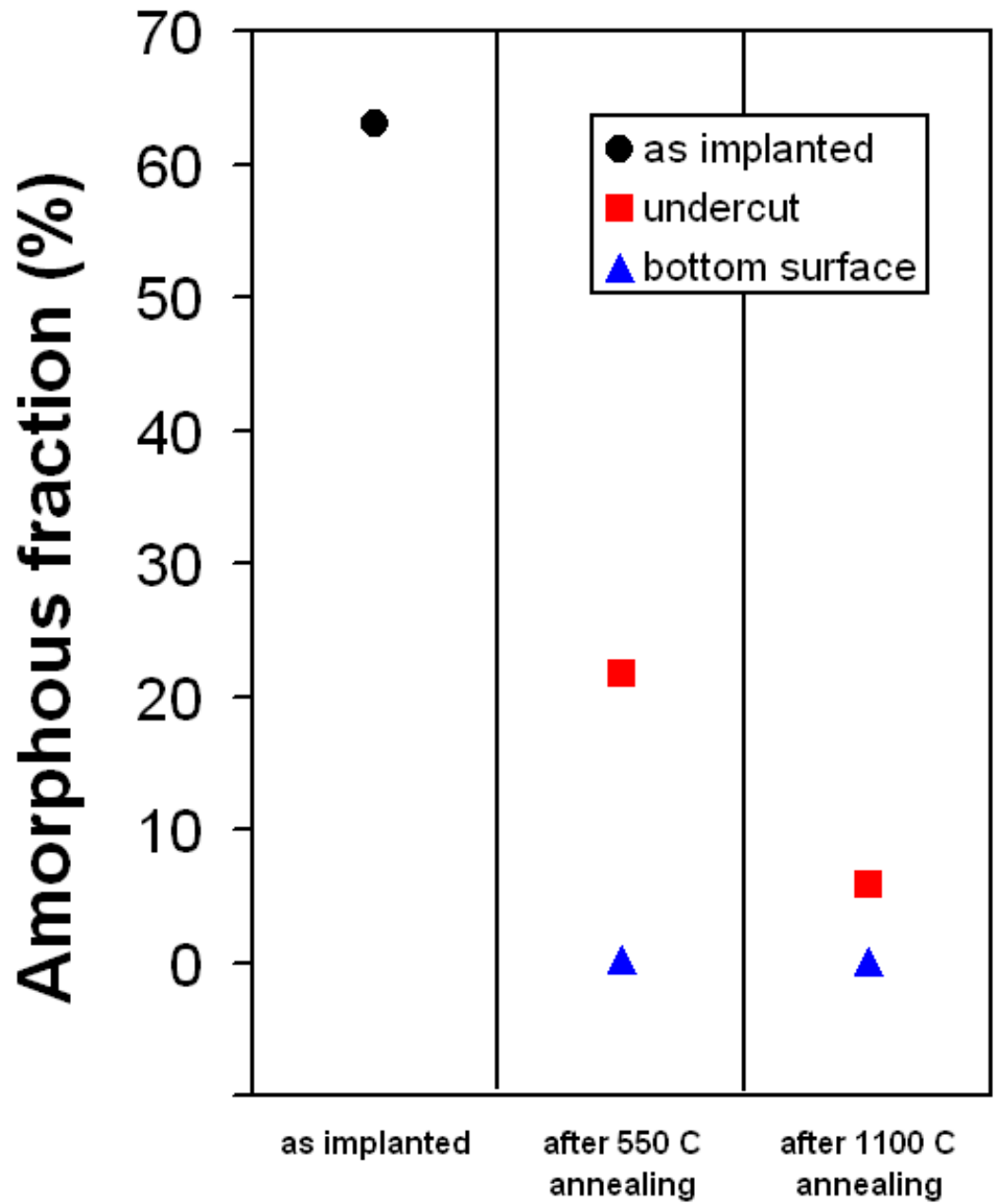

**Fig. 7**

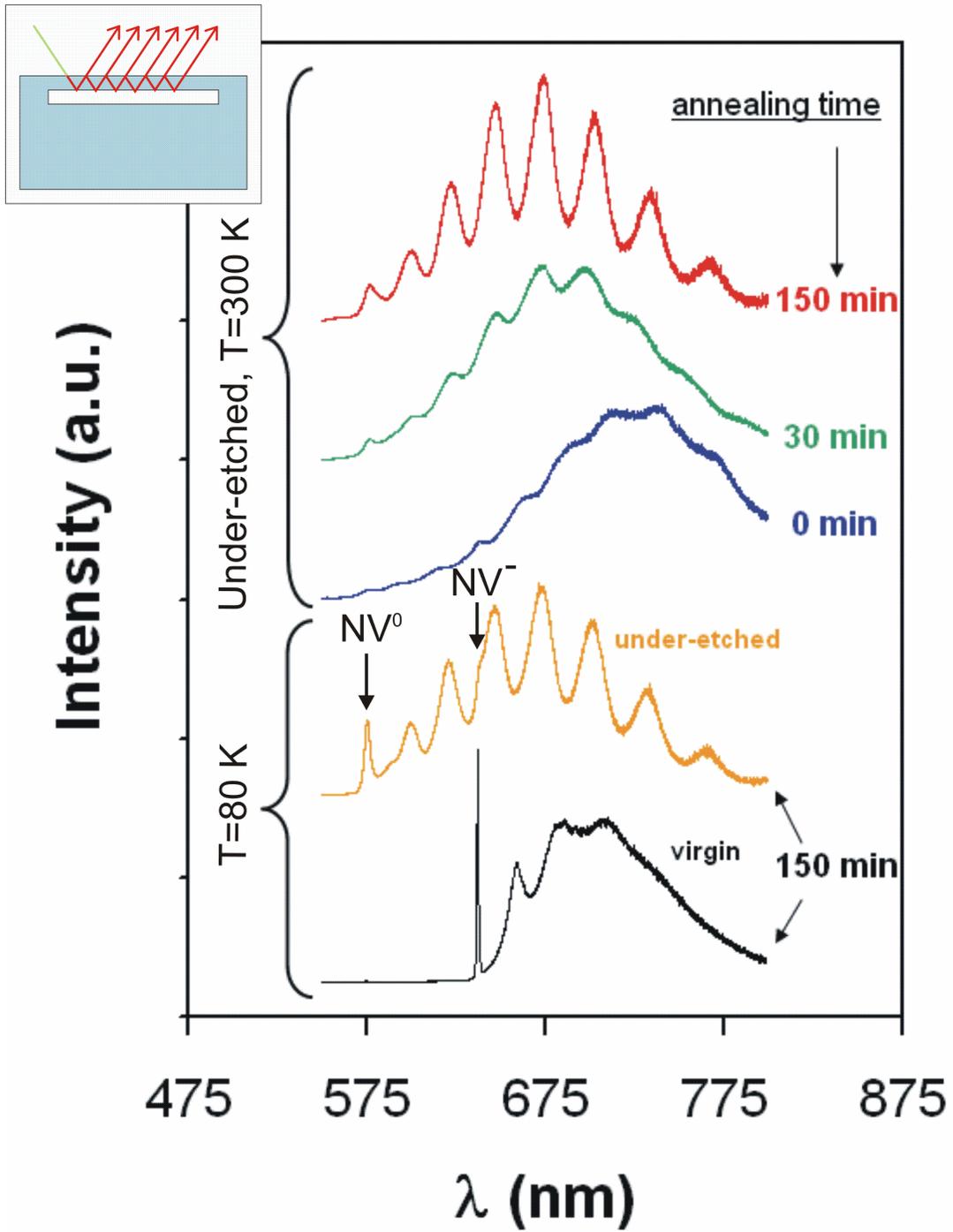